\documentclass[12pt]{article}
\usepackage{amsmath,amssymb}
\title{{\LARGE GAUGE FIELD THEORY OF}\\
       {\LARGE HORIZONTAL SU(2)$\times$U(1) SYMMETRY}\\
\medskip
 - {\large DOUBLET PLUS SINGLET SCHEME } -}
\date{}

\newcommand{\disp}{\displaystyle}
\newcommand{\pr}{\prime}

\newcommand{\pd}{\partial}

\def\bt{\bar{t}}

\def\th{\mbox {$\theta$}}
\def\vth{\vartheta}
\def\btau{\mbox {$\bar{\tau}$}}
\def\be{\mbox {$\bar{e}$}}
\def\bg{\mbox {$\bar{g}$}}
\def\bm{\mbox {$\bar{m}$}}
\def\bM{\mbox {$\bar{M}$}}
\def\bY{\mbox {$\bar{Y}$}}
\def\lam{\mbox {$\lambda$}}
\def\Lam{\mbox {$\Lambda$}}
\def\brlam{\mbox {$\bar{\lambda}$}}
\def\brLam{\mbox {$\bar{\Lambda}$}}
\def\tphi{\mbox {$\tilde{\phi}$}}
\def\vphi{\mbox {${\varphi}$}}
\def\tvphi{\mbox {$\tilde{{\varphi}}$}}
\def\bth{\mbox {$\bar{\theta}$}}
\def\vth{\mbox {$\vartheta$}}
\def\bA{\bar{A}}

\def\bv{\bar{v}}
\def\bW{\bar{W}}

\def\bZ{\bar{Z}}
\def\cD{{\cal D}}
\def\cL{{\cal L}}
\def\cM{{\cal M}}
\def\cY{{\cal Y}}
\def\dlam{\dot{\lambda}}
\def\half{{1 \over 2}}

\def\sqr2{\sqrt{2}}

\begin{document}
\begin{flushright}
MISC-2010-09
\end{flushright}
\vspace{0.6cm}

\begin{center}
{\Large{\bf 
GAUGE FIELD THEORY OF\\ \medskip HORIZONTAL SU(2)$\times$U(1) SYMMETRY}}\\
\bigskip
 - {\large DOUBLET PLUS SINGLET SCHEME } -
\end{center}
\vspace{1cm}

\begin{center}
Ikuo S. Sogami\footnote{E-mail: sogami@cc.kyoto-su.ac.jp }
\end{center}
\begin{center}
{\it Maskawa Institute for Science and Culture, Kyoto Sangyo University,}\\
{\it Kyoto, 603-8555, Japan}
\end{center}

\vspace{1cm}

\abstract{
Gauge field theory of a horizontal symmetry of the group
G$_{\rm H}$ = SU(2)$_{\rm H}\times$U(1) is developed so as to generalize
the standard model of particle physics. All fermion and scalar fields are
assumed to belong to doublets and singlets of the group in high energy
regime. Mass matrices with four texture zeros of Dirac and Majorana types
are systematically derived. In addition to seven scalar particles, the
theory predicts existence of one peculiar vector particle which seems
to play important roles in astrophysics and particle physics.
}
%%%%%%%%%%%%%%%%%%%%%%%%%%%%%%%%%%%%%%%%%%%%%%%%%%%%%%%%%%%%%%%%%%%%%%%%%%%%%%%%%%%%%%%%%%%%%%%%%
%%                                                                                             %%
%%                                     Section 1                                               %%
%%                                                                                             %%
%%%%%%%%%%%%%%%%%%%%%%%%%%%%%%%%%%%%%%%%%%%%%%%%%%%%%%%%%%%%%%%%%%%%%%%%%%%%%%%%%%%%%%%%%%%%%%%%%
\section{Introduction}
To generalize the standard model of particle physics, we develop
a gauge field theory of a horizontal (H) symmetry of the group
G$_{\rm H}$ = SU(2)$_{\rm H}\times$U(1). Above a high energy scale
$\brLam$ which is much higher than the electroweak (EW) scale
$\Lam$, fundamental fermions, quarks and leptons, are postulated to
form doublets and singlets of the group G$_{\rm H}$. Classification of
the fundamental fermions into chiral sectors consisting of EW doublets
($f = q, \ell$) and singlets ($f = u, d; \nu, e$) is assumed to hold also
in the high energy regime.

Breakdown of the symmetry at the scale $\brLam$ ($\Lam$) necessitates
Higgs fields of doublet and singlet of the H symmetry which belong to
the singlets (doublets) of the EW symmetry. The doublet and singlet
composition of the EW and H symmetries for both of the fermion and
scalar fields simplifies the formalism and enables us to reduce the
number of Yukawa coupling constants. In this theory, mass matrices with
four texture zeros of Dirac and Majorana types are systematically derived,
and unphysical modes of bosonic fields are excluded by properly adjusting
values of parameters in the Higgs potentials.

For the sake of distinction, we use the symbols
$\{\,\tau_1,\,\tau_2,\,\tau_3\,\}$ and $Y$ for the isospin and
hypercharge of the EW symmetry, and the symbols
$\{\,\btau_1,\,\btau_2,\,\btau_3\,\}$ and $\bY$ for the
{\lq\lq}isospin{\rq\rq} and {\lq\lq}hypercharge{\rq\rq}
of the H symmetry. The color degrees of freedom are not
specified for simplicity. We introduce a symbol $\bt$ to
indicate the operation of transposition for degrees of the H symmetry. 

\section{Doublet and singlet composition}
The gauge fields of the EW symmetry, $A^{(2)j}_\mu$ ($j=1,\,2,\,3)$
and $A^{(1)}_\mu$, interact to the currents of EW-isospin $\tau_j$
and hypercharge $Y$ with coupling constants $g_2$ and $g_1$.
In contrast, we introduce gauge fields of the H symmetry, $\bA^{(2)j}_\mu$ ($j=1,\,2,\,3$) and $\bA^{(1)}_\mu$, which interact vectorially to
the currents generated by H-isospin $\btau_j$ and H-hypercharge $\bY$
with coupling constants $\bg_2$ and $\bg_1$.

In high-energy region $(>\brLam)$, fundamental fermions in sector
$f\, (= q, u, d; \ell, \nu, e)$ are postulated to belong to the doublet and
singlet of the group G$_{\rm H}$ as follows:
\begin{equation}
 \psi_d^{\,f} = {}^{\bt}\left(
    \begin{array}{cc}
      \psi_{1}^{\,f}, & \psi_{2}^{\,f}
    \end{array}
    \right) ,\quad
    \psi_s^{\,f} = \left(\psi_{3}^f\right),
   \label{doubletsinglet}
\end{equation}
whose components are either the EW chiral doublets as
\begin{equation}
\begin{array}{cc}
  \psi_{i}^{\,q} = 
  \left(
    \begin{array}{c}
      \psi_{i}^{\,u}\\
      \noalign{\vskip 0.2cm}
      \psi_{i}^{\,d}
    \end{array}
  \right)_L ; 
 & 
  \psi_{i}^{\,\ell} = 
  \left(
    \begin{array}{c}
      \psi_{i}^{\,\nu}\\
      \noalign{\vskip 0.2cm}
      \psi_{i}^{\,e}
    \end{array}
  \right)_L
\end{array}
\end{equation}
or the EW chiral singlets as
\begin{equation}
\begin{array}{cccc}
   \left(\psi_{i}^u\right)_R,& (\psi_{i}^d)_R\,;&
   \ \left(\psi_{i}^{\nu}\right)_R,& \left(\psi_{i}^e\right)_R .
\end{array}
\end{equation}
In the low-energy region ($\leq\Lam$), the doublet $\psi_d^{\,f}$ and the
singlet $\psi_s^{\,f}$ turn out to constitute, respectively, main components
of the first and second generations and the third generation of fundamental
fermions.

To properly break the H and EW symmetries, two types of H multiplets of
Higgs fields are presumed to exist. For the H symmetry breaking around
the high-energy scale $\brLam$, a set of doublet and singlet of Higgs
fields are introduced as
\begin{equation}
 \phi_d = 
  {}^{\bt}\left(
    \begin{array}{cc}
      \phi_{1}, &
      \phi_{2}
    \end{array}
  \right) ,\quad
    \phi_s = \left(\,\phi_{3}\,\right),
\end{equation}
where complex fields $\phi_1$ and $\phi_2$ and a real field $\phi_3$ belong
to EW singlets. These scalar fields do not couple with the
fundamental fermion fields in (\ref{doubletsinglet}) except for the right-handed
neutrino fields $\psi_{d, s}^{\nu}$. It is this character of $\phi_a$ that protects the fundamental fermion fields from acquiring Dirac masses of the scale $\brLam$.
A dual doublet of $\phi_d$ is defined by $\tphi_d=i\btau_2\phi_d^\ast$.

To form the Yukawa interaction and break its symmetry at the scale $\Lam$,
a set of H doublet and singlet consisting of three EW doublets must exist as
\begin{equation}
 \vphi_d = 
  {}^{\bt}\left(
    \begin{array}{cc}
      \vphi_{1}, &
      \vphi_{2}
    \end{array}
  \right) =
  {}^{\bt}\left(
    \begin{array}{cc}
  \left(
    \begin{array}{c}
      \vphi_{1}^{\,+}\\
      \noalign{\vskip 0.2cm}
      \vphi_{1}^{\,0}
    \end{array}
  \right), &
  \left(
    \begin{array}{c}
      \vphi_{2}^{\,+}\\
      \noalign{\vskip 0.2cm}
      \vphi_{2}^{\,0}
    \end{array}
  \right)
    \end{array}
  \right)
  ,\quad
 \vphi_s = 
  \left(
    \begin{array}{c}
      \vphi_{3}^{\,+}\\
      \noalign{\vskip 0.2cm}
      \vphi_{3}^{\,0}
    \end{array}
  \right),
\end{equation}
which, respectively, have dual multiplets
$\tvphi_d = (i\btau_2)(i\tau_2)\vphi_d^{\ast}$
and $\tvphi_s = i\tau_2\vphi_s^\ast$.

\section{Lagrangian density}
The Lagrangian density for the fermion and scalar interactions consists of
the Yukawa and Majorana parts.
The density of the Yukawa interaction, $\cL_{\rm Y}^f$, consists of the
{EW}$\times${H} invariants of the multiplets $\psi_a$ and $\vphi_a$ $(a=d, s)$
as follows:
\begin{equation}
 \cL_{\rm Y}^{\,f}
  = \cY_1^f\,\bar{\psi}_d^{\,f^{\pr}}\,\tvphi_s \psi_d^{\,f}
   + \cY_2^f\,\bar{\psi}_d^{\,f^{\pr}}\,\tvphi_d \psi_s^{\,f}
   + \cY_3^f\,\bar{\psi}_s^{\,f^{\pr}}{}^{\bt}\tvphi_d i\btau_2\psi_d^{\,f}
   + \cY_4^f\,\bar{\psi}_s^{\,f^{\pr}}\tvphi_s\psi_s^{\,f}
   + {\rm h.c.}
 \label{YukawaUp}
\end{equation}
for the EW up-\,sectors $(f^\pr = q,\,f=u)$ and $(f^\pr = \ell,\,f=\nu)$, and
\begin{equation}
 \cL_{\rm Y}^{\,f}
  =\cY_1^f\,\bar{\psi}_d^{\,f^{\pr}}\vphi_s \psi_d^{\,f}
   +\cY_2^f\,\bar{\psi}_d^{\,f^{\pr}}\vphi_d \psi_s^{\,f}
   +\cY_3^f\,\bar{\psi}_s^{\,f^{\pr}}{}^{\bt}\vphi_di\btau_2\psi_d^{\,f}
   +\cY_4^f\,\bar{\psi}_s^{\,f^{\pr}}\vphi_s\psi_s^{\,f}
   + {\rm h.c.}
 \label{YukawaDown}
\end{equation}
for the EW down\,-sectors $(f^\pr = q,\,f=d)$ and $(f^\pr = \ell,\,f=e)$.
Four unknown complex coupling constants $\cY_{fi}$ $(i=1,\,\cdots,\,4)$
exist in each sector. The Lagrangian density for the Majorana interaction,
$\cL_{\rm M}$, is given by
\begin{equation}
\cL_{\rm M}
= \disp{\bg}\left(\overline{\psi_d^{\,\nu c}}\,\btau_2\phi_d\psi_s^{\,\nu}
 + \overline{\psi_s^{\,\nu c}}\,{}^{\bt}\phi_d\btau_2\psi_d^{\,\nu}\right)
 +\disp{\bM}_d\overline{\psi_d^{\,\nu c}}\,\btau_2\,\psi_d^{\,\nu}
  +\disp{\bM}_s\overline{\psi_s^{\,\nu c}}\,\psi_s^{\,\nu},
 \label{MajoranaInt}
\end{equation}
where $\psi_a^{\,\nu c}$ are the charge conjugate fields, and $\bg$
and $\bM_a$ $(a=d, s)$ are the Majorana coupling constant and masses.

The Lagrangian density for the scalar fields, $\cL_{\rm scalar}$, takes
the form
\begin{equation}
   \cL_{\rm scalar} = \disp\sum_{a=d,s}(\cD^\mu\phi_a)^\dag(\cD_\mu\phi_a)
 + \disp\sum_{a=d,s}(\cD^\mu\vphi_a)^\dag(\cD_\mu\vphi_a)
 - V_{\rm T}(\phi,\,\vphi) ,
\end{equation}
where $V_{\rm T}(\phi,\,\vphi)$ is the total Higgs potential including all
Higgs fields. The covariant derivatives $\cD_\mu$ for the scalar multiplets
$\phi_a$ and $\vphi_a$ are given, respectively,
by~\footnote{The H-hypercharge of $\phi_d$ is chosen to be 1 by adjusting
the value of the coupling constant $\bg_2$.}
\begin{equation}
 \cD_\mu\phi_d = \left(\pd_\mu - i\bg_2\,\bA^{(2)j}_\mu\,\half\,\btau_j
        - i\bg_1\,\bA^{(1)}_\mu\,\half \right)\phi_d\;,
   \label{covariantderivforbvphid}
\end{equation}
\begin{equation}
 \cD_\mu\phi_s = \pd_\mu\phi_s\;,
   \label{covariantderivforbvphis}
\end{equation}
\begin{equation}
\hspace{-0.1cm}
 \cD_\mu\vphi_d\!=\!\left(\pd_\mu - ig_2\,A^{(2)j}_\mu\,\half\,\tau_j
                              - ig_1\,A^{(1)}_\mu\half
              - i\bg_2\,\bA^{(2)j}_\mu\,\half\,\btau_j
              - i\bg_1\,\bA^{(1)}_\mu\,\half\bY_{\tiny\vphi_d}\right)\vphi_d,
   \label{covariantderivforvphid}
\end{equation}
and
\begin{equation}
 \cD_\mu\vphi_s = \left(\pd_\mu - ig_2\,A^{(2)j}_\mu\,\half\,\tau_j
                              - ig_1\,A^{(1)}_\mu\half
              - i\bg_1\,\bA^{(1)}_\mu\,\half\bY_{\tiny\vphi_s} \right)\vphi_s\;.
   \label{covariantderivforvphid}
\end{equation}

The total Higgs potential of the multiplets $\phi_a$ and $\vphi_a$, $V_{\rm T}(\phi,\,\vphi)$, can be separated into the sum of three parts as follows:
\begin{equation}
  V_{\rm T}(\phi,\,\vphi) = V_1(\phi) + V_2(\vphi) + V_3(\vphi;\,\phi).
\end{equation}
The potential $V_1(\phi)$ of the self-interactions of the multiplets
$\phi_a$ is given by
\begin{equation}
 V_1({\phi}) = -\bm_d^2\,\phi_d^\dag\phi_d
                   -\bm_s^2\,\phi_s^2
                + \half\,\brlam_d\left(\phi_d^\dag\phi_d\right)^2
                + \half\,\brlam_s\,\phi_s^4
                + \brlam_{ds}\left(\phi_d^\dag\phi_d\right)\,\phi_s^2 ,
 \label{bphipot}
\end{equation}
where $\brlam_s, \brlam_d$ and $\brlam_{ds}$ are positive coupling constants
satisfying $\brlam_d\brlam_s>\brlam_{ds}^2$. Using this density, we analyze
the breakdown of the H symmetry around the scale $\brLam$. The potential
$V_2(\vphi)$ of the self-interactions of the multiplets $\vphi_a$ is
expressed as
\begin{eqnarray}
   V_2(\vphi) &=&
       - m_d^2 \vphi_d^\dag\vphi_d - m_s^2 \vphi_s^\dag\vphi_s
       + \half \lam_d(\vphi_d^\dag\vphi_d)^2
       + \half \lam_{d1}|\vphi_d^\dag\tvphi_d|^2
    \nonumber\\
     &&+ \half\lam_{d2}
         \overline{\rm Tr}\left({}^{\bt}\vphi_d^\dag{}^{\bt}\vphi_d\right)^2
       + \half\lam_{d3}
         \overline{\rm Tr}\left({}^{\bt}\vphi_d^\dag{}^{\bt}\vphi_d
         {}^{\bt}\tvphi_d^\dag{}^{\bt}\tvphi_d\right)
       + \half \lam_s(\vphi_s^\dag\vphi_s)^2
\nonumber\\
     && + \lam_{ds}\left(\vphi_d^\dag\vphi_d\right)
           \left(\vphi_s^\dag\vphi_s\right)
         + \lam_{ds1}|\vphi_d^\dag\vphi_s|^2
         + \lam_{ds2}|\vphi_d^\dag\tvphi_s|^2 ,
  \label{phipot}
\end{eqnarray}
where $\overline{\rm Tr}$ means to take the trace operation with respect to
the H-degrees of freedom.
For the potential of mutual interactions between the multiplets $\vphi_a$
and $\phi_a$, we obtain
\begin{eqnarray}
  V_3(\vphi;\,\phi) &=&
           \dlam_1(\phi_d^\dag\phi_d)(\vphi_d^\dag\vphi_d)
         + \dlam_2\,\phi_s^2(\vphi_d^\dag\vphi_d)
         + \dlam_3(\phi_d^\dag\phi_d)(\vphi_s^\dag\vphi_s)
      \nonumber   \\
      \noalign{\vskip 0.2cm}
      &&  + \dlam_4\,\phi_s^2(\vphi_s^\dag\vphi_s)
           + \dlam_5|\tphi_d^\dag\vphi_d|^2
           + \dlam_6|\phi_d^\dag\vphi_d|^2 .
   \label{bphiphipot}
\end{eqnarray}

\section{Symmetry breakdown at high-energy scale $\brLam$}
In the broken phase of the H symmetry around and below the scale $\brLam$,
the doublet and singlet, $\phi_d$ and $\phi_s$, are decomposed into 
the following forms:
\begin{equation}
  \phi_d = {}^{\bt}\left(
                \begin{array}{cc}
                  0\,, \,\  
                 \bv_d + \frac{1}{\sqrt{2}}\xi_d
                \end{array}
               \right),\quad
  \phi_s = \bv_s + \frac{1}{\sqrt{2}}\xi_s,
  \label{phidecomposition}
\end{equation}
where $\bv_d$ and $\bv_s$ are vacuum expectation values (VEVs), and
$\xi_d$ and $\xi_s$ are real component scalar fields. Up to the second order,
the potential $V_1(\phi)$ takes the form
\begin{equation}
  V_1(\phi) = V_1(\bv)
              + \disp \brlam_d\,\bv_d^2\xi_d^2 + \brlam_s\bv_s^2\,\xi_s^2
              + 2\brlam_{ds}\bv_d\bv_s\,\xi_d\xi_s 
            = V_1(\bv)
              + \disp \half m_{\xi_1}^2\xi_1^2 + \half m_{\xi_2}^2\xi_2^2,
\end{equation}
where new real fields $\xi_i$ $(i=1,\,2)$ are introduced by
\begin{equation}
     \xi_d = \cos\bth\,\xi_1 - \sin\bth\,\xi_2,\quad
     \xi_s = \sin\bth\,\xi_1 + \cos\bth\,\xi_2\;.
\end{equation}
The mixing angle $\bth$ is subject to
\begin{equation}
  \tan2\bth = \frac{2\brlam_{ds}\bv_d\bv_s}{\brlam_d\bv_d^2 - \brlam_s\bv_s^2}
\end{equation}
and the mass of the field $\xi_i$ is obtained by
\begin{equation}
   m_{\xi_i}^2 = \brlam_d\bv_d^2 + \brlam_s\bv_s^2
    +(-1)^{i}\sqrt{\left(\brlam_d\bv_d^2-\brlam_s\bv_s^2\right)^2
    + 4\left(\brlam_{ds}\bv_d\bv_s\right)^2}\;.
  \label{massxi12}
\end{equation}

The symmetry breaking at the scale $\brLam$ metamorphoses the gauge fields
$\bA^{(2)j}_\mu$ and $\bA^{(1)}_\mu$ into new fields. Estimation of
the action of the covariant derivative on the scalar doublet at the
stationary state with the VEV $\bv_d$ results in
\begin{equation}
  \left(\cD_\mu\langle\phi_d\rangle\right)^\dag
     \left(\cD^\mu\langle\phi_d\rangle\right)
       = M_{\bW}^2\bW_\mu\bW^\mu + \half M_{\bZ}^2\bZ_\mu\bZ^\mu,
 \label{bAtobWbYbZ}
\end{equation}
where
$\bW_\mu$ is the complex vector field
\begin{eqnarray}
    \bW_\mu = \frac{\bA^{(2)1}_\mu - i\bA^{(2)2}_\mu}{\sqrt{2}}
\end{eqnarray}
with the mass $M_{\bW}^2 = \half\bg^2\bv^2$, and $\bZ_\mu$ is
the neutral vector field
\begin{equation}
      \bZ_\mu = \frac{\bg_2\bA^{(2)3}_\mu - \bg_1\bA^{(1)}_\mu}
      {\sqrt{\bg_2^2+\bg_1^2}}
      = \bA^{(2)3}_\mu\cos\vth - \bA^{(1)}_\mu\sin\vth
\end{equation}
carrying the mass
\begin{equation}
      M_{\bZ}^2 = \half(\bg_2+\bg_1^2)\bv_d^2 = \frac{M_{\bW}^2}{\cos^2\vth}\;.
\end{equation}

There exists another vector field $X_\mu$, being orthogonal to $\bZ_\mu$,
with the configuration
\begin{equation}
      X_\mu = \frac{\bg_1\bA^{(2)3}_\mu + \bg_2\bA^{(1)}_\mu}
      {\sqrt{\bg_2^2+\bg_1^2}}
      = \bA^{(2)3}_\mu\sin\vth + \bA^{(1)}_\mu\cos\vth ,
\end{equation}
which remains massless down to the scale $\Lam$ and makes gauge interaction
to the vector currents of a new charge $\bar{Q}=\half\btau_3 + \half\bY$ of
the H symmetry with the unit of strength, $\be$, defined by
\begin{equation}
     \be = \bg_2\sin\vth = \bg_1\cos\vth\;.
\end{equation}

Substitution of the decompositions of $\phi_a$ in (\ref{phidecomposition})
into (\ref{MajoranaInt}) leads to the effective Lagrangian density of neutrino
species as
\begin{equation}
 \cL_{\rm M}\ \rightarrow\ 
  \cL^{\rm M}_{\cM} = \overline{\Psi^{\nu c}_L}\bar{\cM}_\nu\Psi^\nu_R + \cdots ,
\end{equation}
where $\Psi^{\nu}_{L, R}$ are chiral neutrino triplets in the
interaction mode, $\bar{\cM}_{\nu}$ is the Majorana mass matrix
\begin{equation}
   \bar{\cM}_{\nu} = 
                  \left(\,
                   \begin{array}{ccc}
                    0          &  -i\,\bM_d  & \  -i\bg\bv_d  \\
                  i\,\bM_d     &       0           & \ 0  \\
                  i\bg\bv_d    &       0           & \ \bM_s  \\
                   \end{array}
                  \,\right)\;,
   \label{Majoranamass}
\end{equation}
and the ellipsis means interactions between the neutrinos and scalar
fields $\xi_i$.

\section{Symmetry breakdown at low-energy scale $\Lam$}
To go down to the low-energy region around the scale $\Lam$, the effects
of the renormalization group must properly be taken into account for all
of the physical quantities. In particular, all coupling constants run down
to the values at the scale $\Lam$. For the sake of simplicity, the same
symbols are used here for the quantities including all these effects.

In the broken phase of EW symmetry around the scale $\Lam$, the multiplets
$\vphi_a$ are postulated to take the forms
\begin{equation}
 \vphi_d={}^{\bt}\left(
             \begin{array}{cc} 
             \left(
             \begin{array}{c}
              \zeta_1^{+}\\
              \noalign{\vskip 0.2cm}
                 \zeta_1^0
             \end{array}
             \right), &             
             \left(
             \begin{array}{c}
              \zeta_2^{+}\\
              \noalign{\vskip 0.2cm}
                v_d + \frac{1}{\sqrt{2}}\eta_d
             \end{array}
             \right)
             \end{array}
             \right),\quad
 \vphi_s = \left(
               \begin{array}{c} 
               0\\
               \noalign{\vskip 0.2cm}
               v_s + \frac{1}{\sqrt{2}}\eta_s
               \end{array}
              \right)
    \label{Phidecomposition}
\end{equation}
with VEVs $v_d$ and $v_s$, where $\zeta^{+}_1$, $\zeta^{+}_2$ and $\zeta^{0}_1$
are complex component fields, and $\eta_d$ and $\eta_s$ are real component fields. 
To examine the dynamics around the scale $\Lam$, it is necessary to examine
the sum of the potential $V_2(\vphi)$ and also the potential $V_3(\vphi, \bv)$
which reflects the influence of the H symmetry breakdown at the high-energy scale
$\brLam$. We obtain the stationary conditions as follows:
\begin{equation}
 \begin{array}{ll}
  (\lam_d+\lam_{d2})v_d^2+(\lam_{ds}+\lam_{ds1})v_s^2 &= m_d^{\pr 2}
  \equiv m_d^{2}-\dlam_1\bv_d^2-\dlam_2\bv_s^2-\dlam_6\bv_d^2,\\
 \noalign{\vskip 0.2cm}
  (\lam_{ds}+\lam_{ds1})v_d^2+\lam_{s}v_s^2 &= m_s^{\pr 2}
  \equiv m_s^{2}-\dlam_3\bv_d^2-\dlam_4\bv_s^2 .\\
 \end{array}
\end{equation}
Accordingly, for the phase transition to take place ($v_d, v_s \neq 0$),
reduced quantities $m_d^{\pr 2}$ and $m_s^{\pr 2}$ must be positive.

Around the stationary point, the sum of the Higgs potential is decomposed
with respect to the component scalar fields, up to the second order, as
\begin{equation}
 \begin{array}{ll}
  V_2(\vphi)&+V_3(\vphi;\,\bv_d)
    = V_2(v)+V_3(v;\,\bv_d)\\
   \noalign{\vskip 0.2cm}
   &+ m_{\zeta_1^{+}}^2|\zeta_1^+|^2 + m_{\zeta_2^{+}}^2|\zeta_2^+|^2
   + m_{\zeta_1^{0}}^2|\zeta_1^0|^2 + \half m_1^2\eta_1^2
    + \half m_2^2\eta_2^2 \cdots ,
\end{array}
\end{equation}
where $\eta_i$ $(i=1,\,2)$ are new real fields introduced by
\begin{equation}
     \eta_d = \cos\th\,\eta_1 - \sin\th\,\eta_2,\quad
     \eta_s = \sin\th\,\eta_1 + \cos\th\,\eta_2\;.
\end{equation}
The masses of three complex fields $\zeta^{+}_1$, $\zeta^{+}_2$ and $\zeta^0_1$
are calculated to be
\begin{equation}
\begin{array}{l}
 m_{\zeta_1^{+}}^2=(2\lam_{d1}-\lam_{d2}+\lam_{d3})v_d^2
                   +m_{\zeta_2^{+}}^2
                   +m_{\zeta_1^0}^2,\\
 \noalign{\vskip 0.2cm}
 m_{\zeta_2^{+}}^2=(\lam_{ds2}-\lam_{ds1})v_s^2,\\
 \noalign{\vskip 0.2cm}
 m_{\zeta_1^0}^2=(\dlam_{5}-\dlam_{6})\bv_d^2 .
 \label{MassesComplexScalar}
\end{array}
\end{equation}
The two real fields $\eta_i\,(i=1,\,2)$ have the masses as
\begin{equation}
   m_{\eta_i}^2 = D + S
    +(-1)^{i}\sqrt{\left(D-S\right)^2
    + 4(\lam_{ds}+\lam_{ds1})^2v_d^2v_s^2} ,
 \label{MassesRealScalar}
\end{equation}
and the mixing angle $\bth$ is subjects to
\begin{equation}
  \tan2\th = \frac{2(\lam_{ds}+\lam_{ds1})v_dv_s}{D - S} ,
\end{equation}
where the abbreviations
\begin{equation}
  D = (\lam_{d}+\lam_{d2})v_d^2 - \half(\lam_{ds}+\lam_{ds1})v_s^2,\quad
  S = \lam_sv_s^2
  \label{Abbreviations}
\end{equation}
are used.

Results in (\ref{MassesComplexScalar}), (\ref{MassesRealScalar}) and
(\ref{Abbreviations}) demonstrate that the Higgs coupling constants
must satisfy inequality relations so that all complex and real
scalar fields are in physical modes. For example, the inequality
relations $\lam_{ds2}>\lam_{ds1}$ and $\dlam_{5}>\dlam_{6}$ must
hold to make the masses of the fields $\zeta^{+}_2$ and $\zeta^0_1$
positive-definite. From (\ref{MassesRealScalar}), it is shown that
the real field $\eta_i$ with lighter mass must be identified with
the so-called Higgs particle. Note that stringent conditions on the
flavor changing neutral (charged) currents give strong restrictions
on the Higgs coupling constants.

The symmetry breaking at the scale $\Lam$ changes the gauge fields
$A^{(2)j}_\mu$, $A^{(1)}_\mu$ and $X_\mu$ into massive vector fields
by transferring the four degrees of freedom of the scalar multiplets
$\vphi_a$. To determine configurations of the vector fields, we calculate
the action of the covariant derivative on the scalar multiplets
$\vphi_d$ and $\vphi_s$ at their stationary state with the VEVs of $v_d$
and $v_s$ obtaining
\begin{equation}
 \begin{array}{l}
  \disp\sum_{a=d,s}\left(\cD_\mu\langle\phi_a\rangle\right)^\dag
  \left(\cD^\mu\langle\phi_a\rangle\right)
   = \half g_2^2(v_d^2+v_s^2)W_\mu W^\mu\\
   \noalign{\vskip 0.3cm}
   \quad + \disp\frac{1}{4}v_d^2\left[\frac{g_2}{\cos\th_W}Z_\mu
       +\be(1-\bY_{\tiny \vphi_d})X_\mu\right]
       \left[\frac{g_2}{\cos\th_W}Z^\mu
       +\be(1-\bY_{\tiny \vphi_d})X^\mu\right]\\
   \noalign{\vskip 0.3cm}
   \quad + \disp\frac{1}{4}v_s^2\left[\frac{g_2}{\cos\th_W}Z_\mu
       -\be\bY_{\tiny \vphi_s}X_\mu\right]
       \left[\frac{g_2}{\cos\th_W}Z^\mu
       -\be\bY_{\tiny \vphi_s}X^\mu\right] + \cdots ,
 \end{array}
 \label{WZXA}
\end{equation}
where the charged vector field
\begin{equation}
    W_\mu = \frac{A^{(2)1}_\mu - iA^{(2)2}_\mu}{\sqrt{2}} ,
\end{equation}
and the neutral vector field
\begin{equation}
      Z_\mu = \frac{g_2 A^{(2)3}_\mu - g_1 A^{(1)}_\mu}
      {\sqrt{g_2^2+g_1^2}}
      = A^{(2)3}_\mu\cos\th_W - A^{(1)}_\mu\sin\th_W
\end{equation}
and
\begin{equation}
      A_\mu = \frac{g_1 A^{(2)3}_\mu + g_2 A^{(1)}_\mu}
      {\sqrt{g_2^2+g_1^2}}
      = A^{(2)3}_\mu\sin\th_W + A^{(1)}_\mu\cos\th_W
\end{equation}
are introduced, in exactly the same way with the Weinberg-Salam theory by using
the Weinberg angle $\th_W$ related to the unit $e$ of electromagnetic
interaction as
\begin{equation}
   g_2\sin\th_W = g_1\cos\th_W =  e .
\end{equation}
The ellipsis in (\ref{WZXA}) implies mass corrections to the super-massive
vector fields $\bW_\mu$ and $\bZ_\mu$, and mixing interactions of the field
$\bZ_\mu$ with the fields $Z_\mu$ and $X_\mu$. 

The charged vector field $W_\mu$ possesses the mass
\begin{equation}
      M_{W}^2 = \half g_2^2(v_d^2+v_s^2) .
\end{equation}
The quadratic part of the neutral fields $Z_\mu$ and $X_\mu$
in (\ref{WZXA}) is readily diagonalized provided that
\begin{equation}
   (1-\bY_{\tiny \vphi_d})v_d^2 = \bY_{\tiny \vphi_s}v_s^2 .
   \label{Xcondtion}
\end{equation}
Under this condition, the masses of the fields $Z_\mu$ and $X_\mu$ are
determined to be
\begin{equation}
 M_Z^2 = \half\frac{g_2^2}{\cos^2\th_W}(v_d^2+v_s^2)=\frac{M_W^2}{\cos^2\th_W}
 \label{MassofZ}
\end{equation}
and
\begin{equation}
 M_{X}^2 = \half\frac{\be^2 v_s^2}{v_d^2}(v_d^2+v_s^2)\bY_{\tiny \vphi_s}^2
           = \frac{\be^2 v_s^2}{g_2^2 v_d^2}\bY_{\tiny \vphi_s}^2M_W^2 .
\end{equation}
The mass relation in (\ref{MassofZ}) proves that the firmly-established
experimental criterion for the standard model,
$\rho = M_W^2/M_Z^2\cos^2\th_W = 1$, holds also in the present theory.

Through the breakdown of EW symmetry at the scale $\Lam$, the
fermion fields acquire Dirac type masses. Substitution of the decomposition
of $\vphi_a$ in (\ref{Phidecomposition}) into (\ref{YukawaUp})
and (\ref{YukawaDown}) leads to the effective Lagrangian density
for the fermion fields in the low-energy region as
\begin{equation}
 \cL_{\rm Y}\ \rightarrow\ 
 \cL^{\rm Y}_{\cM}
 = \sum_{f=u,d,\nu,e}\,\bar{\Psi}^{\,f}_{L}\cM_f\Psi^{\,f}_{R}
  + {\rm h.c.} + \cdots ,
\end{equation}
where $\Psi^{\,f}_{L,R}$ are the chiral triplets of $f$-sector in the interaction
mode, and $\cM_f$ are the  Dirac mass matrices. The ellipsis stands for the
interactions of fermion and scalar fields. For the up-\,sectors ($f=u,\,\nu$)
of EW symmetry, we deduce the Dirac mass matrices as follows:
\begin{equation}
   \cM_{f} = \left(\ 
               \begin{array}{ccc}
                \cY_1^{u}v_s  &  0            &  \cY_2^{u}v_d \\
                 0            & \cY_1^{u}v_s  &   0           \\
                 0            & \cY_3^{u}v_d  &  \cY_4^{u}v_s \\
               \end{array}
            \ \right) .
   \label{EWUpDirac}
\end{equation}
Likewise, for the down\,-sectors ($f=d,\,e$) of EW symmetry, we obtain
\begin{equation}
   \cM_{f} = \left(\ 
               \begin{array}{ccc}
                \cY_1^{d}v_s  &   0          &    0          \\
                 0            & \cY_1^{d}v_s &  \cY_2^{d}v_d \\
               -\cY_3^{d}v_d  &   0          &  \cY_4^{d}v_s \\
               \end{array}
            \ \right) .
    \label{EWDownDirac}
\end{equation}
Both matrices which have four texture zeros are characterized by four
independent parameters. It should be recognized that all the parameters
except for two can be set to be real by adjusting phases of the doublets
and singlets of the fermion and scalar fields, $\psi_a^f$ and $\vphi_a$,
in the Yukawa interactions in (\ref{YukawaUp}) and (\ref{YukawaDown}).

\section{Discussion}
Thanks to the unique construction of the present theory in which all
the fermion and scalar fields are presumed to belong to the doublet and
singlet representations of the H and EW symmetries, we have succeeded
systematically to obtain simple mass matrices with four texture zeros
as in (\ref{Majoranamass}), (\ref{EWUpDirac}) and (\ref{EWDownDirac}).
Accordingly, it is tempting to inquire why there exists such a kind of
duality that the symmetry SU(2)$\times$U(1) holds both in the H and EW
degrees of freedom.

For the quark sector, the eigenvalue problems for $\cM_f\cM^{\dag}_f$
$(f=u,\,d)$ which have ten adjustable parameters provide sufficient
information on the mass spectra and flavor mixing matrix. Smallness of
neutrino masses is usually explained by the seesaw mechanism in which
the inverse of the Majorana mass matrix with large components works to
suppress the Dirac matrix. Remark that the determinant of the Majorana
mass matrix in (\ref{Majoranamass}) is calculated to be
$|\bar{\cM}_\nu|=-\bM_d^2\bM_s$. Therefore, the seesaw suppression occurs
exclusively by the Majorana masses $\bM_a$ independently of the VEV
$\bv_d$. This observation reveals that the present scheme might be
reinterpreted to have three energy scales, $\bM_a \gg \brLam \gg \Lam$,
rather than two scales, $\brLam \gg \Lam$.

In this theory, values of the coupling constants in the Higgs potential
must be tuned so that the symmetry breakdowns properly take place and
all bosonic fields acquire positive masses. Furthermore, those coupling
constants must obey much stronger conditions so that the stringent
experimental criteria of the flavor changing neutral (charged) currents
are fulfilled.

In addition to the seven scalar particles related to the fields $\xi_1$,
$\xi_2$; $\zeta_1^+$, $\zeta_2^+$, $\zeta_1^0$, $\eta_1$ and $\eta_2$,
the theory predicts existence of one peculiar particle described by
the vector field $X_\mu$ interacting with current of the charge
$\bar{Q}=\half\btau_3+\bY$ of the horizontal symmetry.
Search for the signals of these particles are expected as possible
targets for the LHC experiment. Through massless and massive stages,
the exotic field $X_\mu$ seems to play important roles in astrophysics
and particle physics.

\end{document}